\documentclass[fleqn,usenatbib]{mnras}

\usepackage{newtxtext,newtxmath}
\usepackage{threeparttable}

\usepackage[T1]{fontenc}

\DeclareRobustCommand{\VAN}[3]{#2}
\let\VANthebibliography\thebibliography
\def\thebibliography{\DeclareRobustCommand{\VAN}[3]{##3}\VANthebibliography}

\usepackage{graphicx}	
\usepackage{amsmath}	

\title[WEAVE observations of the Ring Nebula]{WEAVE imaging spectroscopy of NGC 6720: an iron bar in the Ring}

\author[R. Wesson et al.]{R. Wesson,$^{1,2}$\thanks{E-mail: rw@nebulousresearch.org}
J. E. Drew,$^{2}$
M. J. Barlow,$^{2}$
J. García-Rojas,$^{3,4}$
R. Greimel,$^{5}$
D. Jones,$^{3,4}$
A. Manchado,$^{3,4,6}$
\and
R. A. H. Morris,$^{7}$
A. Zijlstra,$^{8}$
P. J. Storey,$^{2}$
J. A. L. Aguerri,$^{3,4}$
S. R. Berlanas,$^{3,4,9}$
E. Carrasco,$^{10}$
\and
G. B. Dalton,$^{11,12}$
E. Gafton,$^{13}$
R. García-Benito,$^{14}$
A. L. González-Morán,$^{10}$
B. T. Gänsicke,$^{15}$
S. Hughes,$^{16}$
\and
S. Jin,$^{17}$
R. Raddi,$^{18}$
R. Sánchez-Janssen,$^{13}$
E. Schallig,$^{11}$
D. J. B. Smith,$^{19}$
S. C. Trager,$^{17}$
N. A. Walton$^{20}$
\\
$^{1}$ School of Physics and Astronomy, Cardiff University, Queens Buildings, The Parade, Cardiff CF24 3AA, UK\\
$^{2}$ Department of Physics and Astronomy, University College London, Gower Street, London WC1E 6BT, UK\\
$^{3}$ Instituto de Astrofísica de Canarias, 38205 La Laguna, Tenerife, Spain\\
$^{4}$ Departamento de Astrofísica, Universidad de La Laguna, E-38206 La Laguna, Tenerife, Spain\\
$^{5}$ RG Science, Schanzelgasse 17, A-8010 Graz, Austria\\
$^{6}$ Consejo Superior de Investigaciones Científicas (CSIC), E-28014 Madrid, Spain\\
$^{7}$ School of Physics, Bristol University, Tyndall Avenue, Bristol, BS8 1TL, UK\\
$^{8}$ Jodrell Bank Centre for Astrophysics, Department of Physics \& Astronomy, The University of Manchester, Oxford Road, Manchester M13 9PL, UK\\
$^{9}$ Centro de Astrobiología. CSIC-INTA. Campus of the European Space Astronomy Centre (ESAC), E-28692 Villanueva de la Cañada, Madrid, Spain\\
$^{10}$ Instituto Nacional de Astrofísica, Óptica y Electrónica, Luis Enrique Erro 1, C.P. 72840, Tonantzintla, Puebla, México\\
$^{11}$ Department of Physics, University of Oxford, Keble Road, Oxford OX1 3RH, UK\\
$^{12}$ RALSpace, STFC Rutherford Appleton Laboratory, Didcot OX11 0QX, UK\\
$^{13}$ Isaac Newton Group of Telescopes, Apartado 321, 38700 Santa Cruz de la Palma, Tenerife, Spain\\
$^{14}$ Instituto de Astrofísica de Andalucía (IAA-CSIC), P.O. Box 3004, 18080 Granada, Spain\\
$^{15}$ Department of Physics, University of Warwick, Coventry, CV4 7AL, UK\\
$^{16}$ MIT Kavli Institute for Astrophysics and Space Research, 70 Vassar St, Cambridge, MA 02139, United States\\
$^{17}$ Kapteyn Astronomical Institute, University of Groningen, Postbus 800, NL9700 AV Groningen, The Netherlands\\
$^{18}$ Universitat Politècnica de Catalunya, Departament de Física, c/ Esteve Terrades 5, E-08860 Castelldefels, Spain\\
$^{19}$ Centre for Astrophysics Research, University of Hertfordshire, College Lane, Hatfield AL10 9AB, UK\\
$^{20}$ Institute of Astronomy, University of Cambridge, Madingley Rd, Cambridge CB3 0HA, UK
}

\date{Accepted XXX. Received YYY; in original form ZZZ}

\pubyear{\the\year{}}

\begin{document}
\label{firstpage}
\pagerange{\pageref{firstpage}--\pageref{lastpage}}
\maketitle

\begin{abstract}
We present spatially resolved spectroscopic observations of the planetary nebula NGC 6720, the Ring Nebula, taken during the science verification phase of WEAVE, a new instrument mounted on the William Herschel Telescope on La Palma. We use the instrument's Large Integral Field Unit (LIFU) to obtain spectra of the Ring Nebula, covering its entire optically bright inner regions as well as parts of its much fainter outer molecular halo. We report the discovery of emission from [Fe~{\sc v}] and [Fe~{\sc vi}] confined to a narrow ``bar'' extending across the central regions of the nebula. No lines of other elements share this morphology or, at the spectral resolving power used ($R \sim 2500$), the same radial velocity.  The extent to which iron in this bar is depleted is presently unclear; comparison with JWST-detected dust continuum emission suggests that some dust grain destruction may be occurring in the region, but there is currently no observational evidence for the $>$ 50~km\,s$^{-1}$ shock waves or $T > 10^6$~K X-ray emitting gas needed to enable this. Where the bar is located along the line of sight through the nebula, and how it was created, are new puzzles to be solved for this iconic planetary nebula.  
\end{abstract}




\section{Introduction}

Planetary nebulae (PNe) represent an important point in the evolution of low- and intermediate-mass stars, just as their ejecta begin to enrich the interstellar medium. Spectroscopic observations of PNe
are a critical step towards a full understanding of their structure and composition. So far, few have been observed spectroscopically using integral field units capable of capturing their full extent. Here, we present the first such optical panchromatic observations of NGC 6720, the Ring Nebula, obtained using the William Herschel Telescope (WHT) during science verification of the WEAVE Large Integral Field Unit (LIFU). WEAVE is the recently installed WHT Enhanced Area Radial Velocity Explorer \citep{Dalton2016,Jin2024}, and the LIFU is its first-commissioned mode of operation (\citealt{Arnaudova2024}).
This follows on from another recent instance of observations of the Ring Nebula with new instrumentation -- those on board the {\em JWST} \citep{wesson2024, Clark2025, Sahai2025}.  Broad-band {\em JWST} infrared images are used here to complement the new spatially resolved optical spectroscopy of this object.



NGC~6720 has long served as a benchmark object for the study of planetary nebulae, owing to its proximity\footnote{The central star's Gaia EDR3 good-quality parallax \citep{GaiaEDR3} inverts to give a distance of $787^{+29}_{-26}$ pc.}, brightness, and nearly face-on orientation. Its detailed morphology -- including the bright ring, outer halo, and inner cavity -- has been extensively analysed across the electromagnetic spectrum (e.g. \citealt{guerrero1997}, \citealt{odell2013a}, \citealt{odell2013b}, \citealt{odell2013c}, \citealt{fang2018}, \citealt{wesson2024}).
Despite this, several key questions remain about its three-dimensional structure, ionization stratification, and internal kinematics. 

In this letter, we highlight the discovery of a peculiar feature of NGC 6720: we find a linear feature lying across the central star that presently appears to emit only in strongly-ionized iron -- specifically, [Fe~{\sc v}] 4227~\AA{} and some weaker [Fe~{\sc vi}] lines. The detection of a `bar' of highly ionized iron emission 
introduces a previously unreported structural component, with implications for the nebula’s excitation mechanisms and, potentially, for the history of mass loss from the progenitor star.
A full analysis of the whole optical nebula, as observed with WEAVE, will follow in a separate paper (Wesson et al., in preparation).

\section{Observations and Data Processing}
\label{sec:obs}

The Ring Nebula was observed using the WEAVE LIFU mode in May and June 2023. The LIFU has 547 fibres, each 2.6 arcsec in diameter, covering a hexagonal 78$\times$90 arcsec$^2$ field of view with a filling factor of 0.55.  The lower-resolution gratings were used for the observations, delivering a spectral resolution of $R\sim2500$ to both the blue and red arms of the spectrograph. The blue arm covers wavelengths from 3600 to 5955~{\AA}, while the red covers 5800 to 9490~{\AA}.

\begin{figure}
    \includegraphics[width=0.5\textwidth]{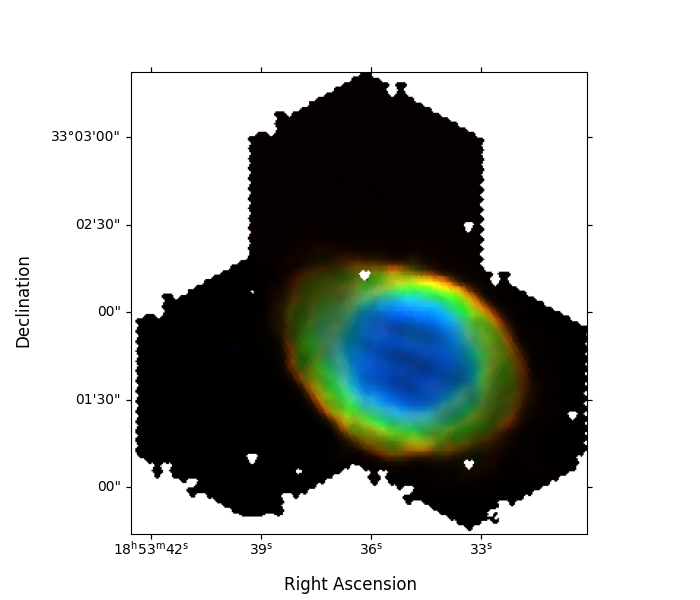}
    \caption{Colour composite image of the Ring Nebula, reconstructed from WEAVE LIFU emission line maps of [O~{\sc i}] 6300~{\AA} (red), H$\beta$ 4861~{\AA} (green), and He~{\sc ii} 4686~{\AA} (blue). Gaps in the reconstructed image arise from LIFU fibres which were not operational at the time of the observations.}
    \label{ring_field}
\end{figure}

Observations were obtained at three pointings, with the field centres set such that the bright ring is completely covered by the mosaicked field (Fig.~\ref{ring_field}).  At each position, the nebula was observed twice, with each observation consisting of three 17-minute exposures, dithered in a three-point pattern to ensure complete spatial coverage of the field of view. A list of the observations used in this study is given in Table~\ref{observationstable}.



\begin{table*}
\begin{tabular}{lllllll}
\hline
OB ID & Date & Arm & RA & Dec & Airmass & Seeing \newline (arcsec) \\
\hline
4876  & 2023-05-19 & Blue & 18:53:33.23 & +33:07:11.79 & 1.06 & 1.02 \\
      & 2023-05-19 & Red  & 18:53:33.23 & +33:07:11.79 & 1.06 & 1.02 \\
4875  & 2023-05-19 & Blue & 18:53:36.03 & +33:08:17.19 & 1.01 & 0.81 \\
      & 2023-05-19 & Red  & 18:53:36.03 & +33:08:17.19 & 1.01 & 0.81 \\
4874  & 2023-05-19 & Blue & 18:53:36.03 & +33:08:17.19 & 1.01 & 0.81 \\
      & 2023-05-19 & Red  & 18:53:36.03 & +33:08:17.19 & 1.01 & 0.81 \\
4872  & 2023-05-20 & Blue & 18:53:39.13 & +33:07:14.44 & 1.15 & 0.65 \\
      & 2023-05-20 & Red  & 18:53:39.13 & +33:07:14.44 & 1.15 & 0.65 \\
4873  & 2023-05-20 & Blue & 18:53:39.13 & +33:07:14.44 & 1.07 & 0.60 \\
      & 2023-05-20 & Red  & 18:53:39.13 & +33:07:14.44 & 1.07 & 0.60 \\
5233  & 2023-06-24 & Blue & 18:53:33.23 & +33:07:11.79 & 1.03 & 1.07 \\
      & 2023-06-24 & Red  & 18:53:33.23 & +33:07:11.79 & 1.03 & 0.86 \\
\hline
\end{tabular}
\caption{List of WEAVE science verification observations of the Ring Nebula}
\label{observationstable}
\end{table*}


Data extraction and wavelength calibration were carried out using the WEAVE core processing system, operated by CASU in Cambridge. This results in the production of a data cube for each OB, in which 
the three dithers are resampled onto an equatorial coordinate grid to form a cube, with wavelengths corrected to the heliocentric reference frame. The two cubes at each pointing were then combined. We have opted to work with the data without pipeline sky subtraction, preferring to remove sky lines later as part of the customised nebular analysis. 
More detail on the LIFU processing may be found in \cite{Arnaudova2024}. Finally, the three merged cubes were mosaicked to form a single data cube,\footnote{using the python package {\sc reproject} (\url{https://reproject.readthedocs.io/})} from which Fig.~\ref{ring_field} and other images in this paper have been produced.    
As these data were obtained in the early stages of WEAVE's operation, there was no systematic collection of calibration observations and the instrument configuration was subject to change.  This resulted in there being no relevant contemporaneous flux calibration data. However, the central star of the Ring Nebula, a very hot white dwarf,\footnote{\textit{Gaia} DR3 2090486618786534784: $G$=15.64, $BP$=15.59, $RP$=16.13} is bright and well-exposed, permitting its use as a flux calibrator.  We compare the observed central-star spectral energy distribution (SED) with a theoretical SED \citep{Csukai2025}, which fits the observed SED presented by \citet{Sahai2025}, to obtain a corrected sensitivity function that is then applied across the nebula.  The theoretical SED uses a non-LTE TMAP model, optimized for PN central stars, with $T_{\rm eff} = 130\,$kK and $\log g = 7$ \citep{Werner2003}, reddened to $E(B-V)=0.05$ \citep{Sahai2025, Csukai2025} using the extinction law of \citet{Fitzpatrick1999}. 

\section{Emission line maps}

We measured both sky and nebular emission line fluxes in the data cube using {\sc alfa}, the Automated Line Fitting Algorithm (\citealt{wesson2016}). This publicly available code measures emission line fluxes by first subtracting an estimate of the nebular continuum. Then, on the continuum-subtracted spectrum, the code generates a population of 50 synthetic spectra with arbitrary emission line fluxes. It applies a genetic algorithm to optimise the parameters of the emission lines to achieve a good fit to the observed spectrum.

Emission lines to be fitted by {\sc alfa} are taken from an input catalogue of wavelengths. When operating on non-sky-subtracted spectra, it first fits sky emission lines using a catalogue of sky line wavelengths from \citet{2003A&A...407.1157H}, fixing their line-of-sight velocity to zero. After fitting and subtracting sky emission, fits to the nebular emission lines are computed in a second pass. We used an input catalogue containing 168 emission lines to fit the full data cube. A full analysis of the spatially resolved emission-line spectra will be published in a forthcoming paper (Wesson et al., in prep).

\section{The iron bar}

\begin{figure*}
    \includegraphics[width=0.95\textwidth]{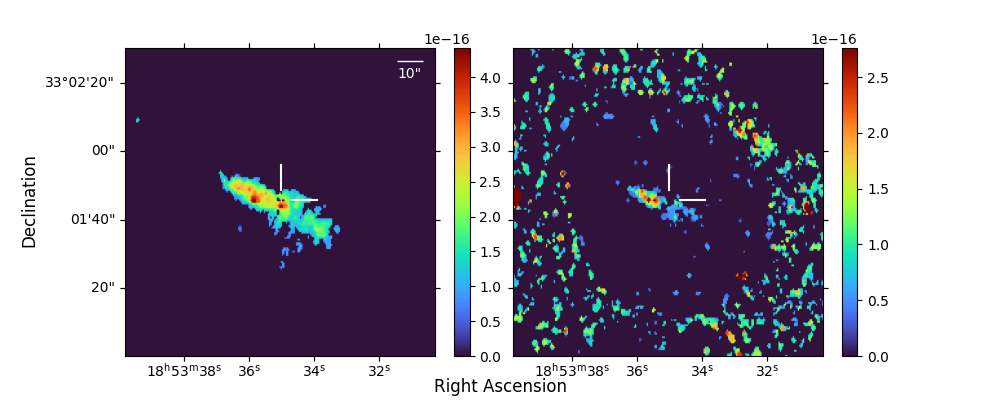}
    \caption{(left) [Fe~{\sc v}] 4227~{\AA} emission line map, (right) summed emission of [Fe~{\sc vi}] emission lines at 5147, 5177, 5425 and 5678~{\AA}. White crosshairs indicate the position of the central star. Apparent emission in the outer regions of the right-hand panel is an artefact of the low signal-to-noise ratio of the [Fe~{\sc vi}] lines.}
    \label{feimage}
\end{figure*}

\begin{figure*}
    \includegraphics[width=\textwidth]{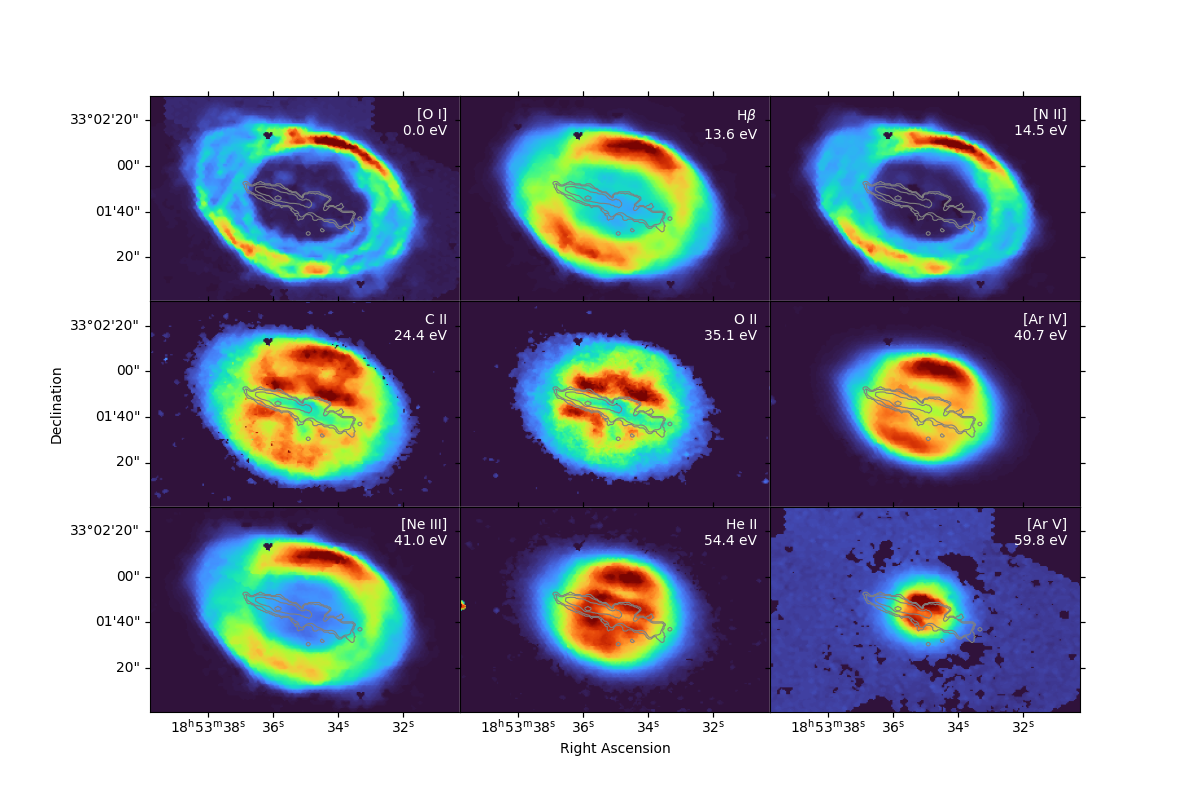}
    \caption{Selected WEAVE emission-line maps, ordered by the minimum photon energy required to create the species being traced: the relevant ion is the detected ion in the case of collisionally-excited lines, and the recombining ion in the case of recombination lines. The emitting species and the associated creation energy are indicated on each map. Contours derived from the [Fe~{\sc v}] map are overlaid in each panel. [Fe~{\sc v}] is produced by collisional excitation of Fe$^{4+}$ (54.8 eV ionization potential). All images are displayed on a linear surface brightness scale with cuts at intensity percentiles of 0.1\% and 99.9\%.}
    \label{weave_comparisons}
\end{figure*}

From the spatially resolved emission line flux measurements, we constructed maps of the nebula in each emission line. An unexpected result revealed by these line maps is the presence of a narrow `bar' of emission at 4227~{\AA} (Fig.~\ref{feimage}). We identify this line as  [Fe~{\sc v}] 4227.20~{\AA}. The `bar' has a length of about 50 arcsec and lies along the major axis of the nebula, at a position angle measured east from north of about 70$^\circ$. The morphology of this region is clearly distinct from that traced by other emission lines (Fig.~\ref{weave_comparisons}). Previous spectroscopic studies of the Ring which employed long-slit spectroscopy (e.g. \citealt{garnett2001}) would not have been able to detect this emission line if the slit was not aligned with the bar. This highlights the significant advantages of spatially-resolved spectroscopy using instruments such as WEAVE to study extended objects.

\begin{figure*}
    \includegraphics[width=0.9\textwidth]{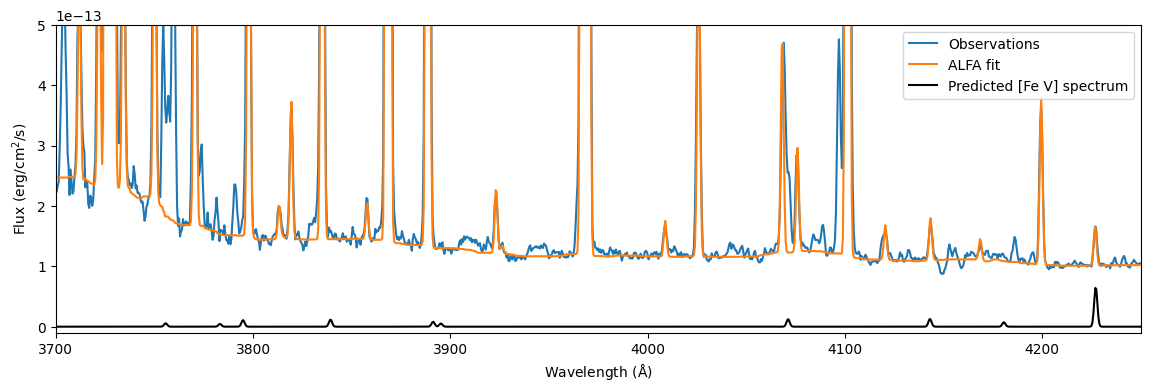}
    \includegraphics[width=0.9\textwidth]{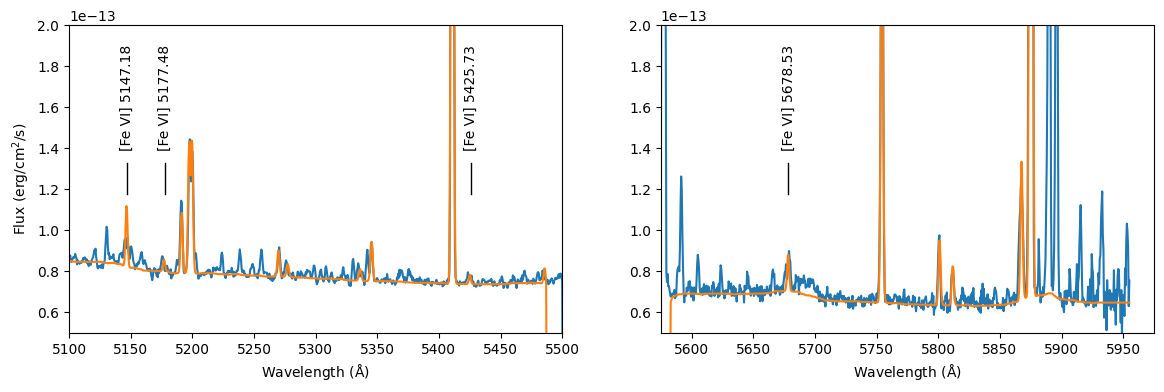}
    \caption{(top) Spectrum of the bar region, showing the clear detection of [Fe~{\sc v}] 4227~{\AA} near the upper wavelength limit of the top panel. The spectrum of [Fe~{\sc v}] predicted for a temperature of 10\,000~K and an electron density of 1000~cm$^{-3}$ is shown in black. [Fe~{\sc v}] 4227~{\AA} is by far the strongest [Fe~{\sc v}] line in the wavelength region observed, and a number of the already much fainter transitions lie close to other strong emission lines. (bottom) Detections of [Fe~{\sc vi}] lines.}
    \label{fespectrum}
\end{figure*}

Although several other lines of [Fe~{\sc v}] are theoretically detectable in optical spectra, they are all weaker than the 4227~{\AA} line, with the next strongest lines predicted to have fluxes of only $\sim$0.2$\times$ the flux of the 4227~{\AA} line (as shown in Fig.~\ref{fespectrum}). Furthermore, many of them lie close to much brighter nebular emission lines. To confirm the identification as [Fe~{\sc v}], we searched for other lines of highly ionized Fe. We detected several [Fe~{\sc vi}] lines (Fig.~\ref{fespectrum}, bottom panels), but no [Fe~{\sc vii}] lines. The [Fe~{\sc vi}] lines give abundances that are broadly consistent with each other (Section~\ref{sec:abundances}), and also fall within the same `bar' as the [Fe~{\sc v}] line (Fig.~\ref{feimage}), thus confirming the identification of the bar's lines as iron.

Fig.~\ref{weave_comparisons} shows a selection of WEAVE emission line maps, arranged in order of the ionization potentials (IPs) of the species being traced, with contours derived from the [Fe~{\sc v}] 4227~\AA\ image superposed. To ionize Fe$^{3+}$ to Fe$^{4+}$ requires an IP of 54.8~eV. To further ionize to Fe$^{5+}$ requires an IP of 75.0~eV, while Fe$^{5+}$ has an IP of 100~eV. So together, the two iron ions detected require photons with energies of 54.8--100~eV to form. 
The distributions of the recombination lines C~{\sc ii} 4267~\AA\ (from C$^{2+}$, IP range = 24.4--47.9~eV) and O~{\sc ii} 4649~\AA\ (IP range 35.1--54.9~eV) are anti-correlated with that of [Fe~{\sc v}], with gaps, mid-nebula, just where the [Fe~{\sc v}] emission is located.
Ar$^{4+}$ ions span an IP range of 59.8--75.4~eV, similar to that of the Fe$^{4+}$ ions, yet the [Ar~{\sc v}] 7005~\AA\ image in Fig.~3, which is centrally concentrated as expected for such a highly ionized species, shows a gap in the middle, at the position of the [Fe~{\sc v}] 4227~\AA\ emission. The He~{\sc ii} 4686~\AA\ image (IP range 54.4-$\infty$~eV) also does not resemble the [Fe~{\sc v}] and [Fe~{\sc vi}] images. We conclude that degree of ionization alone does not determine 
the observed distributions of the observed Fe$^{4+}$ and Fe$^{5+}$ emission.

\subsection{Physical conditions}
\label{sec:physicalconditions}

To investigate the nature of the iron `bar', we extracted a spectrum from a rectangular region of our mosaicked data cube, capturing the iron emission.  This was centred on the central star, with dimensions 18$\times$6.4 arcsec$^2$ and a position angle (measured east from north) of $72^\circ$. We determine the physical conditions and chemical abundances within the bar region using the code {\sc neat} (\citealt{wesson2012}). These are given in Table~\ref{physicalconditions}. We measure the extinction using the four brightest hydrogen Balmer lines. The ratios of H$\alpha$/H$\beta$, H$\gamma$/H$\beta$ and H$\delta$/H$\beta$ are all close to their expected Case B values,\footnote{As defined by \citet{baker1938}, in Case B, the nebular gas is optically thick to Ly$\alpha$ but optically thin in all other transitions} pointing to low interstellar extinction. Previous studies have also found low interstellar extinction; \citet{Sahai2025} find A$_v$=0.15~mag, corresponding to a logarithmic extinction at H$\beta$, c(H$\beta$), of $\sim$0.07. We proceed assuming that c(H$\beta$)=0.0, but if we were to adopt a value of c(H$\beta$)=0.1, the effect on our results would be small: the electron temperature $T_{\rm e}$([O~{\sc iii}]) would be $\sim$100K higher, and our derived Fe abundances would be about 2--3 per cent lower.

Electron densities and temperatures are measured from a number of standard diagnostic lines. The results are shown in Table~\ref{physicalconditions}. The electron densities from different diagnostics are in good agreement with each other, and we take 460~cm$^{-3}$, the error-weighted mean of the three measurements, to use in abundance determinations.

The temperatures calculated from [O~{\sc ii}] and [S~{\sc ii}] line ratios are much higher than other diagnostics. In both cases, the diagnostic lines are split between the red and blue spectral arms, and so their ratios may be more subject to systematic uncertainties. Additionally, in the latter case, the diagnostic lines at 4068 and 4076~{\AA} are blended with several O~{\sc ii} recombination lines, making the derived temperature unreliable. The other diagnostics are in much better agreement. The average of the four reliably determined temperature diagnostics, weighted by their estimated uncertainties, is 10\,300~K, but for calculating abundances, we use the [O~{\sc iii}] line value only, as it is measured from well-detected lines which are all in the same spectrograph arm, and recombination of O$^{3+}$ to O$^{2+}$, which could enhance the flux of the $\lambda$4363 line, should be less significant than recombination to singly ionized species.

\begin{table}

\caption{Electron densities and temperatures measured in the bar region of the Ring Nebula. Atomic data references can be found in \citet{wesson2012}. 
\vspace{-2mm}\newline $^a$ These diagnostics are affected by line blends and recombination excitation of otherwise collisionally excited lines.  }
\centering
\begin{tabular}{llllll}
\hline
\multicolumn{2}{l}{Density (cm$^{-3}$)} \\
\hline
{}[O~{\sc ii}]  &  480 $\pm$ 180 \\
{}[S~{\sc ii}]  &  380 $\pm$ 30 \\
{}[Ar~{\sc iv}]  &  480 $\pm$ 100 \\
\hline
\multicolumn{2}{l}{Temperature (K)} \\
\hline
{}[O~{\sc ii}]$^a$  &  17400 $\pm$ 1500 \\
{}[S~{\sc ii}]$^a$  &  19500 $\pm$ 5100 \\
{}[N~{\sc ii}]  &  10100 $\pm$ 100 \\
{}[O~{\sc iii}]  &  11300 $\pm$ 100 \\
{}[Ar~{\sc iii}]  &  9100 $\pm$ 610 \\
{}[S~{\sc iii}]  &  8500 $\pm$ 260 \\
\hline
\end{tabular}
\label{physicalconditions}
\end{table}

\subsection{Iron abundance}
\label{sec:abundances}

\begin{table}
\caption{Iron line intensities and ionic abundances (expressed by number) in the bar region. The observed H$\beta$ line flux in the bar region is (4.96$\pm$0.09)$\times$10$^{-11}$~erg\,cm$^{-2}$\,s$^{-1}$. The solar Fe/H abundance is taken from \citet{lodders2025}.}
\begin{tabular}{lllll}
\hline
Species & Wavelength ({\AA}) & I$_\lambda$ (I$_{H\beta}$=100) & X$^{i+}$/H$^+$ \\
\hline
Fe$^{4+}$ & 4227.20 & 0.25 $\pm$ 0.02 & (7.3 $\pm$ 0.6)$\times$10$^{-8}$ \\
Fe$^{5+}$ & 5147.18 & 0.13 $\pm$ 0.03 & (5.7 $\pm$ 1.4)$\times$10$^{-8}$ \\
Fe$^{5+}$ & 5177.48 & 0.02 $\pm$ 0.01 & (1.2 $\pm$ 0.5)$\times$10$^{-8}$ \\
Fe$^{5+}$ & 5425.73 & 0.02 $\pm$ 0.01 & (2.0 $\pm$ 0.8)$\times$10$^{-8}$ \\
Fe$^{5+}$ & 5678.53 & 0.08 $\pm$ 0.03 & (7.1 $\pm$ 2.5)$\times$10$^{-8}$ \\
Fe$^{5+}$ & average & & (5.4 $\pm$ 2.5)$\times$10$^{-8}$ \\
\hline
Fe$^{4+}$+Fe$^{5+}$ & & & (1.3 $\pm$ 0.3)$\times$10$^{-7}$ \\
Solar Fe/H & & & 3.2$\times$10$^{-5}$ \\
\hline
\end{tabular}

\label{ironabundances}
\end{table}

Table~\ref{ironabundances} presents a list of the iron lines detected in the extracted `bar' spectrum and the abundances derived from them. We calculate abundances using an electron temperature of 11\,300~K and a density of 460~cm$^{-3}$, as determined in Section~\ref{sec:physicalconditions} and set out in Table~\ref{physicalconditions}. In the absence of diagnostic ratios directly pertaining to the iron bar, this is the necessary choice at this time.   For [Fe~{\sc v}], we use collision strengths from \citet{ballance2007} and transition probabilities from \citet{nahar2000}. For [Fe~{\sc vi}], we use collision strengths from \citet{ballance2008} and transition probabilities from \citet{chen2000}.

The abundances we measure for Fe$^{4+}$ and Fe$^{5+}$ are 7.3$\times$10$^{-8}$ and 5.4$\times$10$^{-8}$, respectively, giving an Fe/H abundance of $>$1.3$\times$10$^{-7}$ from these two ions. This is a factor of $\sim$250 below the solar value of 3.24$\times$10$^{-5}$ (\citealt{lodders2025}). This depletion factor represents an upper limit, due to (i) the unobserved lower ionization stages, and (ii) the likelihood that the hydrogen line emitting volume extends further along the line of sight than the iron-emitting volume.

In NGC~6720's bright ring, we detect the [Fe~{\sc iii}] line at 4658.10~{\AA}. This lies very close to C~{\sc iv} 4658.64~{\AA}. Maps of this line show two spatial components, with emission interior to the bright ring having a similar spatial pattern to He~{\sc ii} emission, and emission in the bright ring resembling low-ionization emission lines. Attributing the emission interior to the bright ring to [C~{\sc iv}] and emission in the bright ring to [Fe~{\sc iii}], we derive Fe$^{2+}$/H$^+$=(3.3 $\pm$ 0.3)$\times$10$^{-8}$, a factor of $\sim$1000 below the solar Fe/H abundance. The lower depletion in the bar region relative to the bright ring may suggest that iron in the bar region has been released back into the gas phase by dust destruction. The total mass of iron ions detected in the Ring Nebula's iron bar is 8.5$\times10^{26}$~g (56 per cent Fe$^{4+}$; 44 per cent Fe$^{5+}$), equivalent to 0.14 of an Earth mass.

\subsection{Kinematics}

\begin{figure*}
    \includegraphics[width=0.48\textwidth]{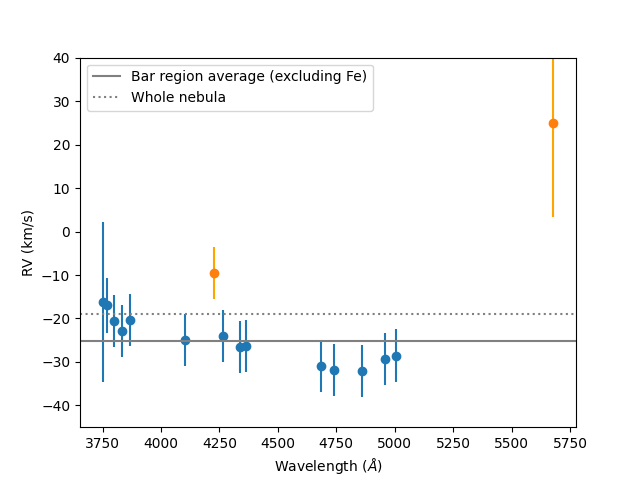}
    \includegraphics[width=0.48\textwidth]{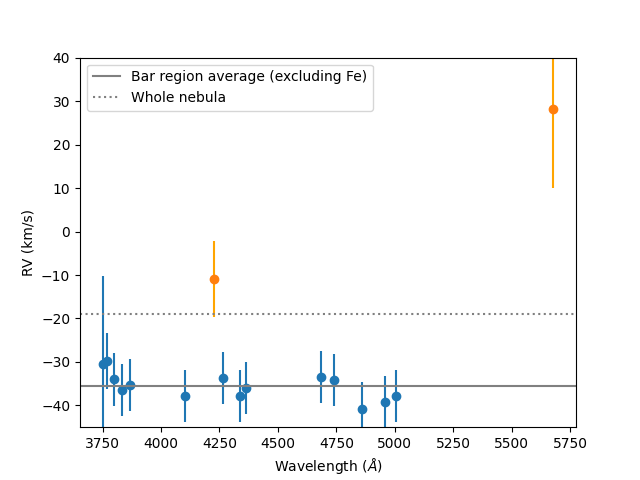}
    \caption{Heliocentric radial velocities measured from individual spectral lines in the blue spectral region, for the bar regions east (left panel) and west (right) of the central star. The [Fe~{\sc v}] 4227~\AA\ and [Fe~{\sc vi}] 5679~\AA\ lines (plotted in orange) stand out as redshifted with respect to the median of the other lines.}
    \label{radialvelocities}
\end{figure*}

The `bar' of Fe emission is brightest on the east side of the central star, but is also clearly present on the west side. Kinematic information may constrain the nature of the emission. Although our observations were taken using the LIFU's low resolution mode, we are able to obtain some useful kinematic information.

We fit Gaussian profiles to a number of emission lines across the red and blue spectral arms, both east and west of the central star. If the `bar' represented some kind of collimated bipolar outflow, one would expect to find red-shifted emission on one side of the star and blue-shifted emission on the other, unless it were to lie close to the plane of the sky.

Fig.~\ref{radialvelocities} shows the radial velocities derived by fitting Gaussian profiles to a number of isolated and well-detected emission lines in the blue spectra. Although there is considerable scatter, most emission lines have broadly consistent radial velocities. Two outliers are apparent; these are [Fe~{\sc v}] 4227~{\AA} and [Fe~{\sc vi}] 5678~{\AA}, which are redshifted relative to other emission lines by $\sim$20~km\,s$^{-1}$ and $\sim$50~km\,s$^{-1}$, respectively. While the average velocity of emission lines excluding Fe lines differs by about 10~km\,s$^{-1}$ between the east and west sides of the central star, the Fe line velocities are the same to within their uncertainties on both sides. These observations are not consistent with a scenario in which the most-ionized Fe emission arises from a bipolar outflow.




\subsection{Comparison with JWST imagery}

\begin{figure*}
    \includegraphics[width=\textwidth]{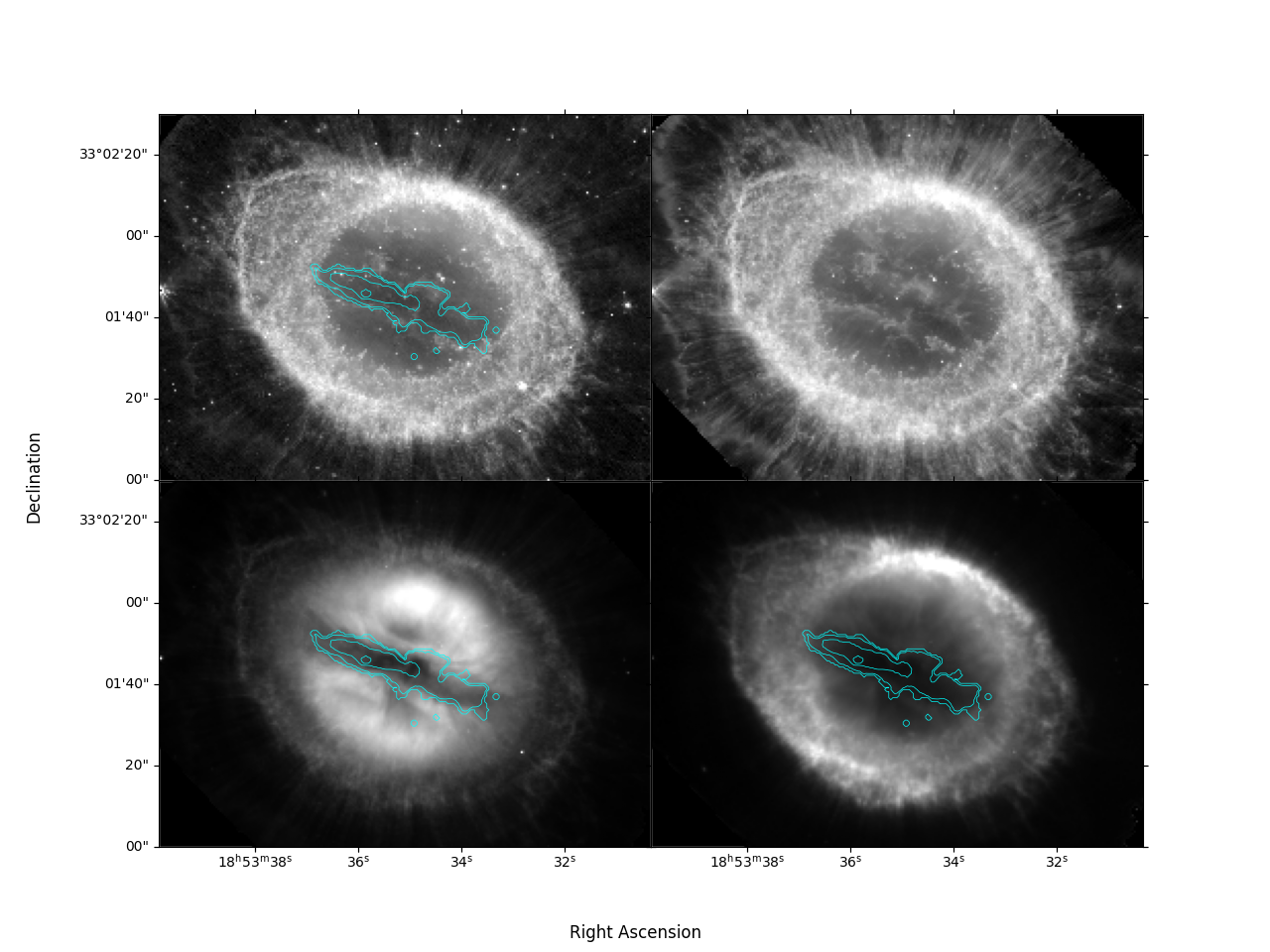}
    \caption{Top row: JWST NIRCam F335W and MIRI F560W. Bottom: MIRI F1000W, MIRI F1800W. Contours derived from the WEAVE map of [Fe~{\sc v}] 4227~{\AA} are displayed, except on the top right panel where they are omitted to better show the H$_2$ emission on either side of the dark lane crossing the centre.}
    \label{jwst_comparison}
\end{figure*}

We compare the WEAVE map of [Fe {\sc v}]~4227~\AA\ emission with broad-band images of the Ring Nebula obtained with JWST and presented in \citet{wesson2024}. Fig.~\ref{jwst_comparison} shows four of the JWST images, representative of the NIR/MIR morphologies present, with [Fe~{\sc v}] emission contours superposed.

The JWST images show two sets of features which correlate with the optical iron emission. Firstly, a number of filters show emission on either side of a dark lane coinciding with the optical iron emission. Filters showing this pattern are F335W and F560W (shown in Fig.~\ref{jwst_comparison}) as well as F770W, F1130W, F1280W and F1500W (not shown). These filters all contain emission from H$_2$ (\citealt{wesson2024}), and so the linear features on either side of the iron bar appear to be H$_2$ emission.

Secondly, in filters F1000W and F1800W (shown in Fig.~\ref{jwst_comparison}) as well as F2100W (not shown), a dark lane spatially coinciding with the optical iron emission is apparent. This is clearest in the F1000W image. \citet{wesson2024} estimated that this filter contains 46\% continuum emission, with emission-line contributions from [S~{\sc iv}] (26\%) and [Ar~{\sc iii}] (14\%). However, these estimates are based on JWST spectroscopy of small regions close to the bright ring. WEAVE image maps show [Ar~{\sc iii}] emission to be much brighter in the main ring than interior to it, and the smooth distribution of emission interior to the ring in the F1000W image also suggests that, in the central region, it predominantly traces continuum emission. The anticorrelation of dust continuum emission with iron emission suggests that the iron is being released into the gas phase as dust is destroyed.



\section{Discussion}

The nature of the iron `bar' in the Ring Nebula is unclear. While fast collimated outflows are commonly seen in planetary nebulae, and the bar appears jet-like in projection, the kinematic information shows that it is not a jet.

The only similar structure in another planetary nebula, to the best of our knowledge, is a recently-reported detection of highly-ionised Fe in spatially-resolved spectroscopic observations of NGC~6818 with MUSE (Monreal-Ibero et al., in prep.\footnote{conference talk available at \url{https://www.youtube.com/watch?v=0JA2FY29zIM&t=2355s}}). However, in that case, the kinematic information appears to be consistent with a jet origin for the highly ionised gas.


The central star of the Ring does not bisect the iron bar, but is offset southeast of bar centre by a few arcsec (Fig.~\ref{feimage}). \citet{wesson2024} find that the central star is itself offset from the centre of the visual cavity by about 2 arcsec in the northwest direction. \citet{Kastner2025} have shown that the star is at the centre of the outer, molecular nebula, and so it is only offset within the visual cavity. The bar, on the other hand, is more nearly on the visual cavity's geometric central axis, and is likely to be associated with the central regions of the nebula.

The (Fe$^{4+}$ + Fe$^{5+}$)/H$^+$ abundance ratio of (1.3$\pm0.3$)$\times10^{-7}$ in Table~3 corresponds to an iron depletion factor of 250 relative to the solar value, consistent with dust formation effects: this indicates that overall, the gas in the nebula derives from the asymptotic giant branch (AGB) wind and not from the central star wind, which does not form dust. However, the hydrogen and other non-iron lines present in the iron bar spectra may very well arise mainly from a
different spatial component, in front of or behind the iron emitting gas. 
Any line components of other species forming in the iron bar region with similar radial velocity offsets of +(15--25)~km~s$^{-1}$ (Fig.~\ref{radialvelocities}) are currently lost against the bright emission from the rest of the nebula. This prevents the true depletion in the iron bar component from being measured. A much higher spectral resolving power of at least 20,000 is needed to separate and detect what would be weak line components of other lines, in the extreme case in which the iron in the bar is undepleted, or nearly so.  In this context, we note that the $R\sim 40\,000$ long-slit spectroscopy presented by \citet{ODell2007} did provide evidence that there is some He~{\sc ii} redshifted emission at comparable velocity (see their fig.~2). 

Radiative transfer models computed using {\sc cloudy} (\citealt{gunasekara2025}) predict that H$\gamma$ 4340~\AA\ in undepleted gas should be approximately as bright as [Fe~{\sc v}] 4227~\AA . Simply multiplying the 4227/H$\beta$ line flux ratio of 2.5$\times10^{-3}$ from Table~3 by the nominal iron depletion factor of 250 gives a similar result, in that a 4227/H$\beta$ line flux ratio of 62\% is implied for undepleted gas.


How a component with a low or even zero gas-phase depletion of iron could be formed in the Ring Nebula is not clear. The sputtering of iron grains or iron-rich silicate grains by shock waves or by very hot gas cannot be invoked: for shock-wave destruction to occur, shock velocities significantly in excess of those materials' sputtering threshold velocity of $\sim$50~km~s$^{-1}$ are needed \citep[][their figs.~5 and 7]{Jones1994}. \citet[][their fig.~8]{Kastner2025} 
have detected compact CO emission components in NGC~6720 at redshifted and blueshifted radial velocities of $\sim$45-50~km~s$^{-1}$ relative to systemic, which seem too low to cause significant grain destruction. 
Similarly, significant thermal sputtering of such grains requires gas temperatures of 10$^6$~K or larger \citep[][their fig.~12]{Tielens1994} -- and yet, no X-ray emission has ever been reported from this low-interstellar-extinction nebula.

At present, there seem to be no obvious explanations that can account for the presence of the narrow `bar' of [Fe~{\sc v}] and [Fe~{\sc vi}] emission seen in our WEAVE spectra to extend across the central regions of the Ring Nebula.  Fresh observations of this newly uncovered feature at much higher spectral resolution seem essential to make progress.


\section*{Acknowledgements}

Based on observations made with the William Herschel Telescope operated on the island of La Palma by the Isaac Newton Group of Telescopes in the Spanish Observatorio del Roque de los Muchachos of the Instituto de Astrofísica de Canarias.

Funding for the WEAVE facility has been provided by UKRI STFC, the University of Oxford, NOVA, NWO, Instituto de Astrofísica de Canarias (IAC), the Isaac Newton Group partners (STFC, NWO, and Spain, led by the IAC), INAF, CNRS-INSU, the Observatoire de Paris, Région Île-de-France, CONACYT through INAOE, the Ministry of Education, Science and Sports of the Republic of Lithuania, Konkoly Observatory (CSFK), Max-Planck-Institut für Astronomie (MPIA Heidelberg), Lund University, the Leibniz Institute for Astrophysics Potsdam (AIP), the Swedish Research Council, the European Commission, and the University of Pennsylvania. The WEAVE Survey Consortium consists of the ING, its three partners, represented by UKRI STFC, NWO, and the IAC, NOVA, INAF, GEPI, INAOE, Vilnius University, FTMC – Center for Physical Sciences and Technology (Vilnius), and individual WEAVE Participants. Please see the relevant footnotes for the WEAVE website\footnote{\url{https://weave-project.atlassian.net/wiki/display/WEAVE}} and for the full list of granting agencies and grants supporting WEAVE\footnote{\url{https://weave-project.atlassian.net/wiki/display/WEAVE/WEAVE+Acknowledgements}}.

RW acknowledges support from STFC Consolidated grant ST/W000830/1.
JG-R and DJ acknowledge support from the Agencia Estatal de Investigaci\'on del Ministerio de Ciencia, Innovaci\'on y Universidades (AEI/MCIU) under grant ``Nebulosas planetarias como clave para comprender la evoluci\'on de estrellas binarias'' and the European Regional Development Fund (ERDF) with reference PID-2022-136653NA-I00 (DOI:10.13039/501100011033). DJ also acknowledges support from AEI/MCIU under grant ``Revolucionando el conocimiento de la evoluci\'on de estrellas poco masivas'' and the the European Union NextGenerationEU/PRTR with reference CNS2023-143910 (DOI:10.13039/501100011033).
AM acknowledges support from the AEI/MCIU, and the ERDF, under grant
PID2023-147325NB-I00/AEI/10.13039/501100011033. This publication is based upon work from COST Action CA21126 - Carbon molecular nanostructures in space (NanoSpace), supported by COST (European Cooperation in Science and Technology).
JALA acknowledges support from the AEI/MCIU under the grant ‘WEAVE: EXPLORING THE COSMIC ORIGINAL SYMPHONY, FROM STARS TO GALAXY CLUSTERS’ and the ERDF, with reference PID2023-153342NB-I00 / 10.13039/501100011033.
RGB acknowledges financial support from the Severo Ochoa grant CEX2021-001131-S funded by MCIN AEI/10.13039/501100011033 and PID2022-141755NB-I00.
DJBS acknowledges support from the United Kingdom’s Science and Technology Facilities Council (STFC) via grant ST/Y001028/1, and from the Leverhulme Trust via Research Project Grant RPG-2025-078.
This project has received funding from the European Research Council (ERC) under the European Union’s Horizon 2020 research and innovation programme (Grant agreement No. 101020057).

We thank Dr Connor Ballance for providing us with the full set of Fe$^{4+}$ collision strengths
calculated by Ballance et al. (2007).

Finally, we thank the anonymous referee for their helpful report.

\section*{Data Availability}

The fully reduced and mosaicked WEAVE data cube will be available in a special data release for the ING community covering all WEAVE Science Verification data, and later as part of the first public, world-wide data release, which will take place two years after the start of full WEAVE Survey operations.



\bibliographystyle{mnras}
\bibliography{references} 



\bsp	
\label{lastpage}
\end{document}